\documentstyle[epsfig]{mn}

\newif\ifAMStwofonts
\AMStwofontstrue

\def\be{\begin{equation}}
\def\ee{\end{equation}}

\def\gtsima{$\; \buildrel > \over \sim \;$}
\def\ltsima{$\; \buildrel < \over \sim \;$}
\def\prosima{$\; \buildrel \propto \over \sim \;$}
\def\gsim{\lower.5ex\hbox{\gtsima}}
\def\lsim{\lower.5ex\hbox{\ltsima}}
\def\simgt{\lower.5ex\hbox{\gtsima}}
\def\simlt{\lower.5ex\hbox{\ltsima}}
\def\simpr{\lower.5ex\hbox{\prosima}}

\def\ie{{\frenchspacing\it i.e. }}
\def\eg{{\frenchspacing\it e.g. }}
\def\etal{{\it et al.~}}

\title[Lyman Continuum Escape]{Lyman Continuum Escape
from Inhomogeneous ISM}
 
\author[B. Ciardi, S. Bianchi \& A. Ferrara]{B. Ciardi$^1$, S. Bianchi$^2$\thanks{Fellow
of the European Community Research and Training Network: \emph{ The Physics of the
Intergalactic Medium}.} and A. Ferrara$^3$\\
$^1$ Max-Planck-Institut f\"ur Astrophysik, Karl-Schwarzschild-Stra\ss e 1, 85748 Garching, Germany\\
$^2$ European Southern Observatory, Karl-Schwarzschild-Stra\ss e 2, 85748 Garching, Germany\\
$^3$ Osservatorio Astrofisico di Arcetri, Largo Enrico Fermi 5, 50125 Firenze, Italy \\} 

\date{September 2001}

\pagerange{\pageref{firstpage}--\pageref{lastpage}}
\pubyear{2001}
\begin{document}

\maketitle
\label{firstpage}

\begin{abstract}
We have studied the effects of gas density inhomogeneities on the 
escape of ionising Lyman continuum (Lyc) photons from Milky Way-type galaxies
via 3D numerical simulations using the Monte Carlo radiative transfer code
{\tt CRASH} (Ciardi \etal 2001). To this aim a comparison
between a smooth Gaussian distribution (GDD) and an inhomogeneous, fractal one
(FDD) has been made with realistic assumptions for the ionising stellar sources
based on available data in the solar neighborhood. 
In both cases the escape fraction $f_{esc}$ 
increases with ionisation rate $\dot{{\cal
N}}_\gamma$ (although for the FDD with a flatter slope) and they
become equal at $\dot{{\cal N}}_\gamma = 2 \times 10^{50}$~s$^{-1}$ where
$f_{esc} = 0.11$. FDD allows  escape fractions of the same order also at
lower $\dot{{\cal N}}_\gamma$, when Lyc photon escape
is sharply suppressed by GDD. Values of the escape fraction as high as 0.6 can be
reached (GDD) for $\dot{{\cal N}}_\gamma \approx 9\times 10^{50}$~s$^{-1}$, 
corresponding to a star
formation rate (SFR) of roughly 2 $M_\odot$~yr$^{-1}$; at this ionising
luminosity the FDD
is less transparent ($f_{esc} \approx 0.28$). If high redshift galaxies
have gas column densities similar to local ones,   
are characterized by such high SFRs and by a predominantly smooth (\ie
turbulence free)
interstellar medium, our results suggest that they should considerably
contribute to - and possibly dominate -  the cosmic UV background.
\end{abstract}


\section{Introduction}

Recent studies suggest that the ionising radiation escaping from galaxies could give
a substantial (possibly dominant) contribution to the ultraviolet background
radiation (UVB) during a large cosmological time span (Giallongo, Fontana \&
Madau 1997; Giroux \& Shull 1997;  Bianchi, Cristiani
\& Kim 2001). Also, at large redshifts, radiation from the first stellar
objects has very likely driven the process of cosmic reionisation (Gnedin
\& Ostriker 1998; Ciardi \etal 2000; Miralda-Escud\'e, Haehnelt \& Rees
2000; Gnedin 2000; Benson \etal 2000; Ciardi \etal
2001). In spite of this extensive body of work on both aspects, their 
predictive power is jeopardized by the persisting (theoretical and
experimental) ignorance on the value of $f_{esc}$, the fraction of 
hydrogen-ionising
photons that escapes from the parent galaxy into the intergalactic medium (IGM).
Obviously, this quantity enters the modeling of the UVB and reionisation 
process and the results depend quite sensibly on the assumptions made about
this poorly constrained parameter.

A promising way to make theoretical progresses on this issue is to 
improve the degree of realism of the modeling and the treatment of 
physical processes to make predictions that can be directly compared with 
available local and intermediate redshift data.

\begin{figure*}
\psfig{figure=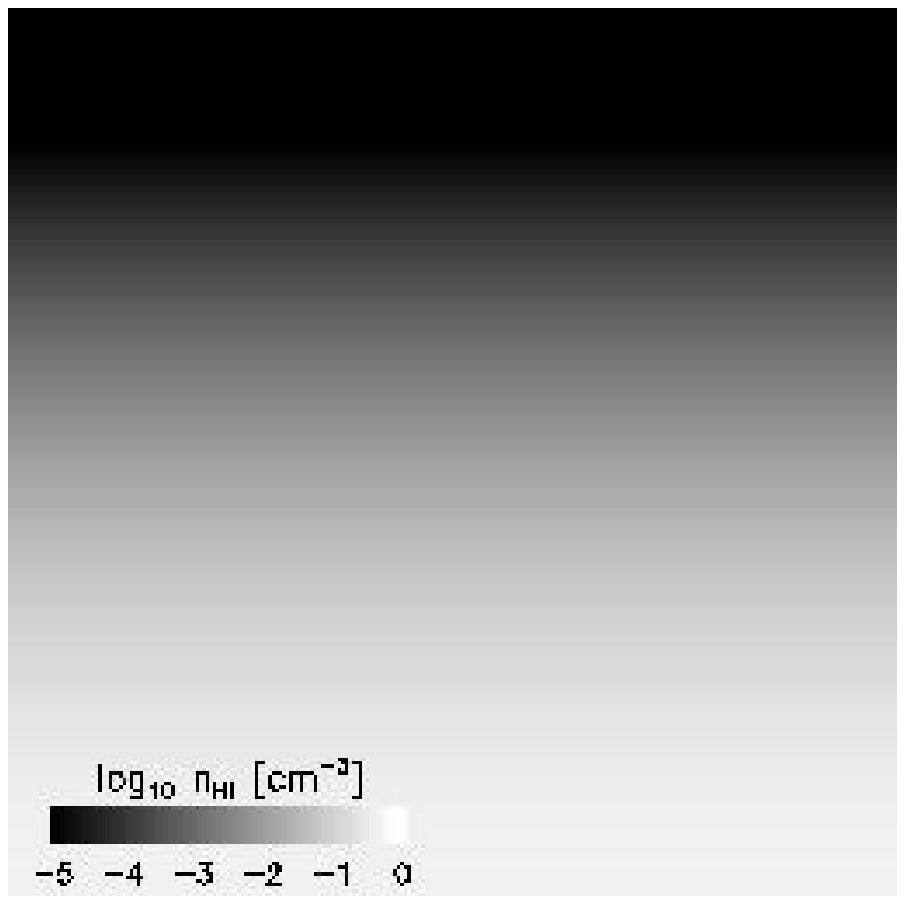,height=8cm}
\psfig{figure=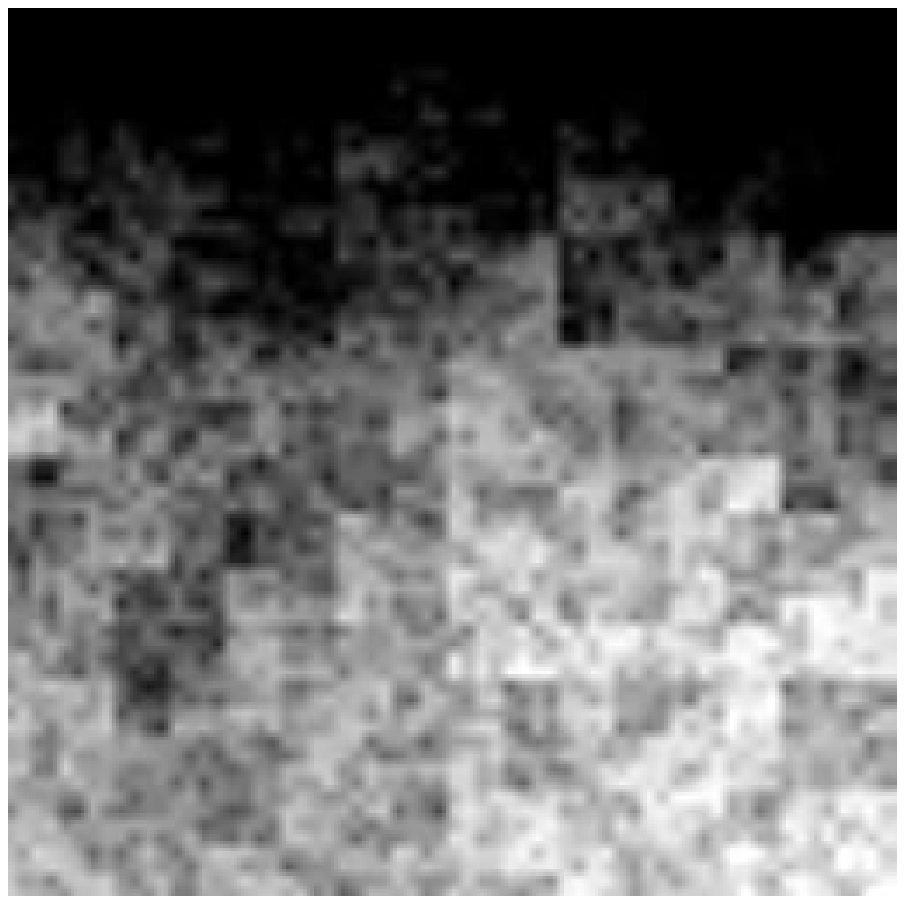,height=8cm}
\caption{\label{fig01}\footnotesize{Slice of the Gaussian
(left panel) and fractal (right panel) density distribution
described in the text, cut through a plane perpendicular to
the Galactic disk. The linear size is 1~kpc.
Images are displayed with the same logarithmic scale.}}
\end{figure*}

Dove \& Shull (1994b) initially tackled the problem of determining 
$f_{esc}$ by assuming smoothly varying H I distributions in
the Galactic disk. They concluded that about 14\%  of Lyman continuum
(Lyc) photons are
able to escape the disk. Dove, Shull \& Ferrara (2000) later improved the
calculation by solving the time-dependent radiation transfer problem 
of stellar radiation through evolving superbubbles. Their main result
is that the shells of the expanding superbubbles
quickly trap ionising photons, so that most of the 
radiation escapes shortly after the formation of the superbubble.
This results in a value of $f_{esc}$ roughly a factor of 2 lower
than obtained by Dove \& Shull (1994b). Additional theoretical works 
(Ricotti \& Shull 2000; Wood \& Loeb 2000) have extended the analysis
to include high-redshift galaxies, for which the escape of Lyc photons
can be modified by geometrical effects (\ie disk vs. spheroidal systems)
and by the higher mean galactic interstellar medium (ISM) density. 
Generally speaking, the escape fraction
decreases as the object virialization redshift or mass become larger;
for a typical 2-$\sigma$ fluctuation in a CDM model at redshift $\approx
10$ these studies derive even lower values ($f_{esc} \simlt 1$\%).

Observationally, a wide range of values for the escape fraction has been
deduced, but it appears that some data favor larger values than expected from theory. 
For example, most of the detections of starburst galaxies with the
{\it Hopkins Ultraviolet Telescope} and the {\it Far Ultraviolet
Spectroscopic Explorer} (Leitherer \etal 1995; Hurwitz,
Jelinsky \& Dixon 1997; Heckman \etal 2001) are consistent with $f_{esc}<10\%$,
although objects have been observed with  $f_{esc}<57\%$ (Hurwitz,             
Jelinsky \& Dixon 1997). However, absorption from undetected interstellar
components could allow the true escape fractions to exceed these upper
limits. Bland-Hawthorn \& Maloney (1999; see Erratum 2001) used optical 
line emission data for the Magellanic Stream to derive $f_{esc} \approx 45$\%. 
Even more puzzling are the recent results (Steidel, Pettini \& Adelberger 2001;
Haenhelt \etal 2001) 
who detected flux beyond the Lyman limit (with significant residual flux
at $\lambda <  912 \AA$) in a composite spectrum of 29 LBGs at $z = 3.4$.
Also, this implies that at these early epochs galaxies were much more transparent
to ionising radiation than at present time, contrary to the trend found by 
the theoretical works mentioned before. 

Can the low values of $f_{esc}$ predicted by theory and 
the high values suggested by recent observations be brought into agreement? In
this paper we try to ascertain if the dichotomy can be removed by considering
the effects of inhomogeneities in the ISM structure and a more realistic 
distribution of massive stars, the primary sources of ionising photons. 
As already pointed out before, essentially all studies (with the partial 
exception of Wood \& Loeb 2000) to date have derived the
escape fraction assuming smoothly varying gas density distributions. 
However, this position is clearly untenable in the light of the large 
number of observations showing that the ISM in the Galaxy and nearby ones
has a hierarchical, very likely fractal, structure (Elmegreen \& Falgarone
1996 and references therein). This view is also supported by theoretical works
(see for example Norman \& Ferrara 1996), which predict a fractal behavior
over a large dynamical range. This type of organization 
naturally arises from turbulent pressure, which plays a crucial role in the
various phases of the ISM (see for example Kulkarni \& Fich [1985] for the cold 
neutral HI component and Reynolds [1985] for the warm ionised medium):
Norman \& Ferrara (1996) showed
that the ISM turbulent pressure is roughly 30 times higher than the thermal
one. Turbulence is pumped into the ISM primarily by multi-SN explosions;
however, reaching the high energy density of present day galaxies requires a
considerable fraction of the Hubble time at
high redshift. Hence the ISM of these primordial systems, differently from
their local counterparts, is likely to be more quiescent and smooth.  

For these reasons, it seems worthwhile to examine a model for the
escape fraction in which a fractal, turbulence dominated ISM is taken
into account. This is what we explore in the rest of the paper. 

\section{Model}                       

In this section we describe the model adopted for the 
Milky Way (MW) gas density and the stellar distribution, 
as well as the source emission properties. 
We use the MW as a reference template, but our results should
hold with good approximation for the entire class of MW-type galaxies, 
as long as the properties of their ISM are similar. 

To study the escape  of (ionising) Lyman continuum (Lyc) photons from the
MW, we have chosen a representative cubic volume of 
1~kpc$^3$ located above the disk midplane and centered at the 
Solar position. This allows us to exploit the high quality data
available for the spatial distribution of the ionising OB stars (\eg Garmany, Conti
\& Chiosi 1982).
Also, the same argument applies to the exquisite detail with which  
the gas density distribution has been derived (Dickey \& Lockman 1990;
Reynolds \& Haffner 2001 and references therein).

Comparison between the soft X-ray background and absorption line  
experiments toward stars near the Sun, have suggested that
we live in a rarefied, hot ISM cavity, usually known as the Local
Bubble, whose average radial extent is $\approx$ 100 pc
(Sanders et al. 1977; Snowden et al. 1990; Sfeir et al. 1999).
As the volume of the Local Bubble is  only 0.2\% of the simulation volume,
we neglect for simplicity the effects of such structure on our results. 

\subsection{ISM Density Distribution}
\label{ismden}

In order to compare the effects of inhomogeneities on the
escape of Lyc photons with previous results, we have first considered 
the commonly adopted Gaussian HI density distribution (GDD): 
\be
n_{HI}(z)=n_0 {\rm exp}(-z^2/2H^2),
\label{gdens}
\ee
where $z$ is the height above the midplane of the disk. 
Such distribution is a solution of hydrostatic equation for the gas in the
gravitational field of a disk galaxy, including a dark matter component.  
The Gaussian distribution that most closely resembles the 
three-component Dickey-Lockman (1990) vertical distribution has 
parameters $n_0=0.367$~cm$^{-3}$ and $H=0.184$~kpc (Dove, Shull \& Ferrara 2000). 
This gives a total neutral hydrogen column density
perpendicular to the disk of $N_{HI}=5.2 \times 10^{20}$~cm$^{-2}$.

Observations of molecular clouds have shown that with
increasing spatial resolution the gas distribution breaks up into
substructures of yet smaller scales. The self-similarity of the 
relations linking some of the molecular cloud properties 
(such as the mass-size relation
and the distribution functions of size and mass) suggests that the 
density distribution very likely has a fractal structure induced by turbulence
(Falgarone, Phillips \& Walker 1991; Elmegreen \& Falgarone 1996;
Elmegreen 1999). 
As a more realistic description, we have thus alternatively 
adopted a fractal model.

A fractal ISM distribution is obtained from hierarchically clustered
points, along the prescriptions given by Elmegreen (1997). In brief,
starting from a point with coordinates $(i_0,j_0)=(0.5,0.5)$,
fractals on the $x-y$ plane were made with $K=6$ hierarchical 
levels by choosing continuous coordinates such that for random
numbers $R_i$ in the interval [0,1], 
$i_1 =i_0+2(R_1 - 0.5)/L^1$ at the first level, $i_2
=i_1 + 2(R_2 - 0.5)/L^2$ at the second level, and so on up to level $K$;
$L = 2$ is a geometric factor for subdivision of one level into the next.
The same procedure was used for $j$. At each level, $N=5$ new positions
were substituted for each position in the previous level, leaving $N^K$
positions after $K$ levels. At each height $z$, a $L^{2K}=64^2$ square
grid was placed around all the points, and the number of points inside each
grid cell was counted. We thus recovered the gas density in each grid cell
constraining its average value to be the same as for the Gaussian
distribution of eq.~\ref{gdens}. The resulting gas distribution has thus
a fractal dimension $D={\rm log}N/{\rm log}L=2.32$, which is consistent
with the experimentally derived value (Elmegreen \& Falgarone 1996). 
This procedure is repeated at each of
the 64 heights that have discrete values in the range $0<z<1$~kpc;
this range represents a good compromise between the need to encompass 
most of the vertical extent of the HI disk and to achieve a sufficient spatial
sampling.  
The clumping factor of this distribution, $C \equiv \langle n^2 \rangle /
\langle n \rangle^2$, calculated as a function of
distance from the midplane, ranges between 4 and 8.

In principle, it would have been possible to use the HI distribution
directly obtained from 21-cm line observations. However, resolving
power limitations make such maps
of little use to study the effects of inhomogeneities. 
For example, the spatial sampling of the Dwingeloo survey 
(Burton \& Hartmann 1994) is $0.5^\circ$ (corresponding to $\approx 0.03$ kpc) 
but the spectral sampling is 1.03 km s$^{-1}$, or 0.1 kpc. This is about 
a factor of 6 lower than the resolution required by our simulation. 
Surveys with higher spatial resolution (1 arcmin) do exist, but
they do not cover a sufficiently large latitude range (the best is probably
the Canadian Galactic Plane Survey which covers $-3.5^\circ$ to $+5.5^\circ$, 
equivalent
to half the simulation box) at roughly the same spectral resolution.
Thus, we have restricted our analysis to the previous two model cases.
A summary of the adopted density distributions in the various simulation
runs is shown in Table~1. For illustration, 
in Fig.~\ref{fig01} we show a slice of the GDD 
(left panel) and the FDD (right panel) cut through a plane perpendicular to 
the Galactic disk.

Measurements of faint optical interstellar emission lines and pulsar
dispersion measures, show    that in the MW 
approximately 1/3 of the HI mass is contained in the Warm Ionized Medium 
(Reynolds 1991a; Reynolds 1993).
To account for the presence of such component, we
have multiplied the GDD and FDD described above by a factor of 1.3. 
We assume that initially the gas has a temperature of  
100~K, as derived  for the Cold Neutral Medium.

\subsection{Stellar Distribution}

\begin{table}
\centerline{Table 1: Simulation Parameters}
\begin{center}
\begin{tabular}{lccc}
\hline
\hline
Run & $n_{HI}^{(1)}$ & $n_\star^{(2)}$ &
$\Sigma_\star$ $^{(3)}$\\
\hline
\hline
A & Gaussian &  Gaussian & 24 \\
B & Fractal  &  Gaussian & 24 \\
C & Gaussian &  Gaussian & 48 \\
D & Fractal  &  Gaussian & 48 \\
E & Fractal  &  Schmidt Law & 24 \\ \hline
\end{tabular}
\end{center}
{$^{(1)}${Density Distribution}}\\
{$^{(2)}${Stellar Distribution}}\\
{$^{(3)}${Stellar Surface Density [kpc$^{-2}$]}}\\
\end{table}

Garmany, Conti \& Chiosi (1982) compiled a catalog of O-type stars in the
Solar neighborhood. The catalog is supposed to be complete to a distance of 
2.5 kpc from the Sun. Within this radius, the surface density of such
component is about $\Sigma_\star=24$ 
stars kpc$^{-2}$. Therefore, we have reproduced the local stellar 
population by randomly extracting positions and photon ionisation rates 
for 24 stars in our 1 kpc$^3$ computational volume as follows.

We have adopted for the stars a vertical Gaussian distribution with a
scaleheight $H_\star=63$~pc, 
as recently derived by Ma\'{\i}z-Apell\'aniz (2001) from a sample 
of O-B5 stars from the Hipparcos catalog.
By randomly sampling this distribution, we have derived the
$z$-coordinate of each star; the $x$ and $y$ coordinates have
been selected randomly. All stars above the 
galactic plane have been included in the simulations. 

As the star formation process should preferentially take place in
the densest regions of the ISM, we have also produced stellar distributions
assuming that the probability of finding a star in a certain region is
proportional to a power of the gas density, $P\propto \rho^{1.5}$, 
\ie a Schmidt-type law (Kennicut 1998).

Finally, we have also produced stellar distributions with a surface density
two times larger than the local one, $\Sigma_\star = 48$ kpc$^{-2}$,
to study the dependence of $f_{esc}$ on a larger range of ionisation rates.
A summary of the adopted stellar distributions is shown in Table~1.

\subsection{Source Emission Properties}

A spectral type and luminosity class has been randomly assigned to each 
star, using the number frequencies of objects of a given type in the
catalog of Garmany, Conti \& Chiosi (1982). Within 2.5 kpc from the Sun, about 50\%
of the stars are of O9.0 - O9.5 types, half of which of luminosity class V. 
We have then used the stellar models of Schaerer \& de Koter (1997) to
derive the rate of ionising photons from the given spectral type and luminosity 
class.

Following this recipe, we have produced 20 different stellar distribution
realizations for each case in Table~1, to get
some semblance of the scatter among different realizations.
The scatter is produced by the random sampling of the  
number, position and ionising photon rate of individual stars 
in the simulation volume. 

For simplicity, we have adopted a single black-body spectrum for each star.
We have assumed a temperature of 40000~K, a mean value for the effective
temperatures of O-type stars ($T_{eff}=30000-50000$ K; Schaerer \&
de Koter, 1997).
The pattern of ionisation does not heavily depend on the detailed shape of 
the spectrum, as long as most of the ionising photons are emitted close
to the HI ionisation limit.

Early B-type stars also produce ionising radiation. Assuming a Salpeter 
Initial Mass Function (IMF) and the ionising photon rate of Schaerer \& de 
Koter (1997), we have estimated that B0-B0.5 stars contribute on average to 
less than 10\% of the total number of ionising photons. We have run a few 
simulations to check the influence of B stars. 
The difference between models with and without B stars are masked 
by the random scatter in the results. Therefore, we did not include B-type 
stars in the stellar distributions. 

\section{Radiative Transfer Simulations}               

\begin{figure*}
\psfig{figure=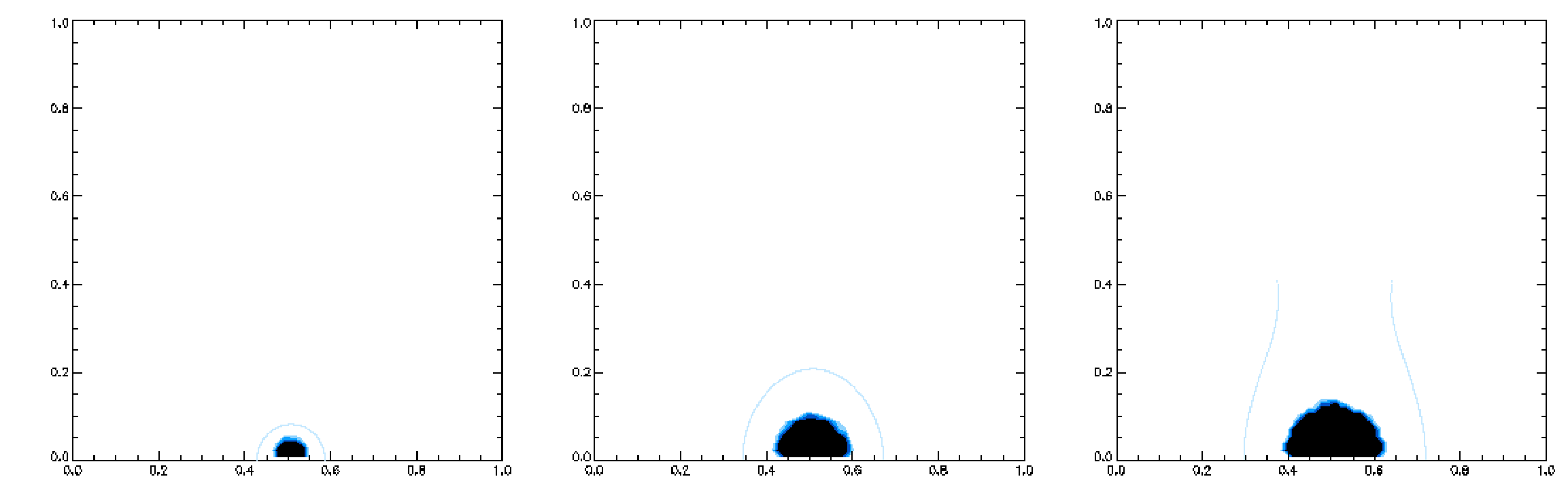,height=5.5cm}
\psfig{figure=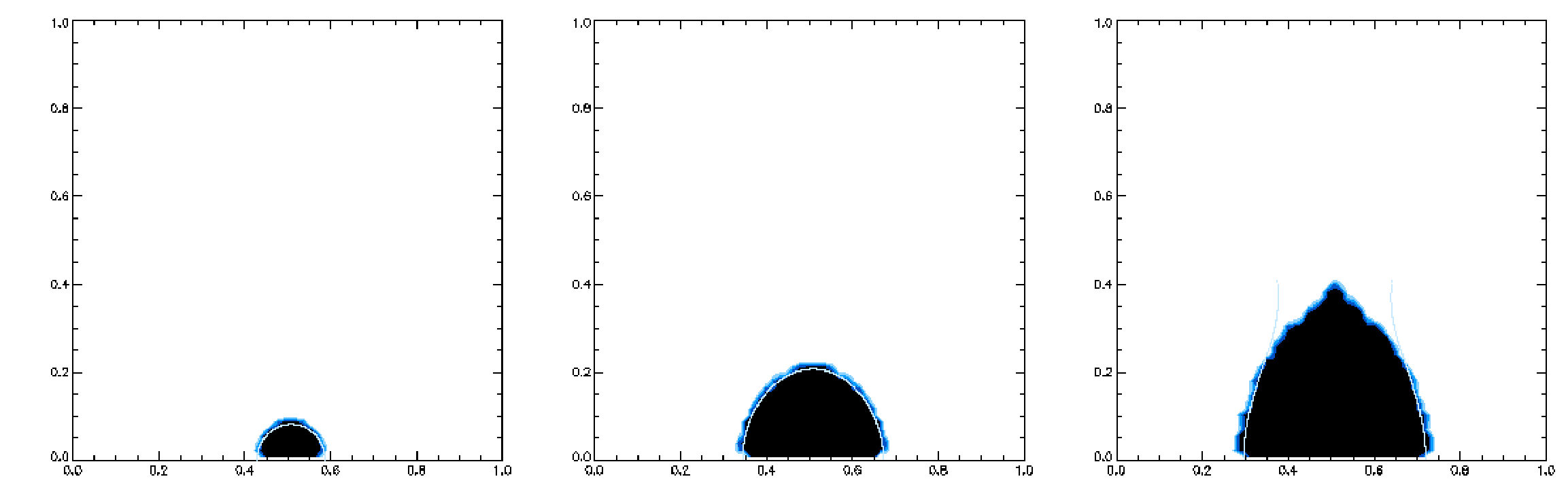,height=5.5cm}
\psfig{figure=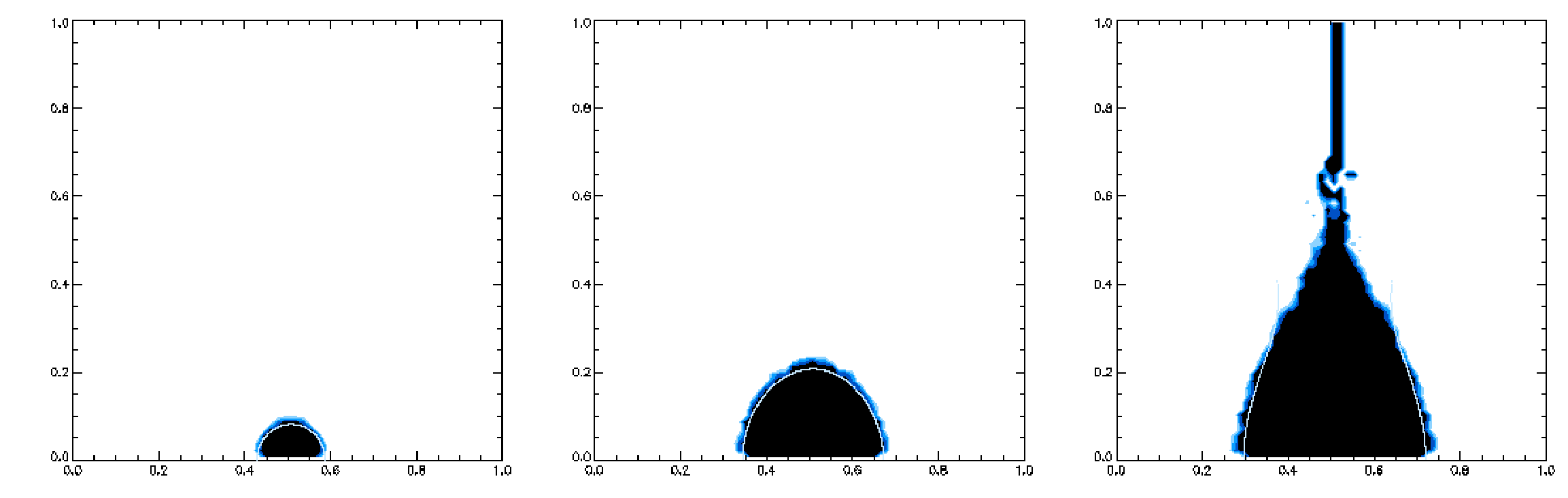,height=5.5cm}
\psfig{figure=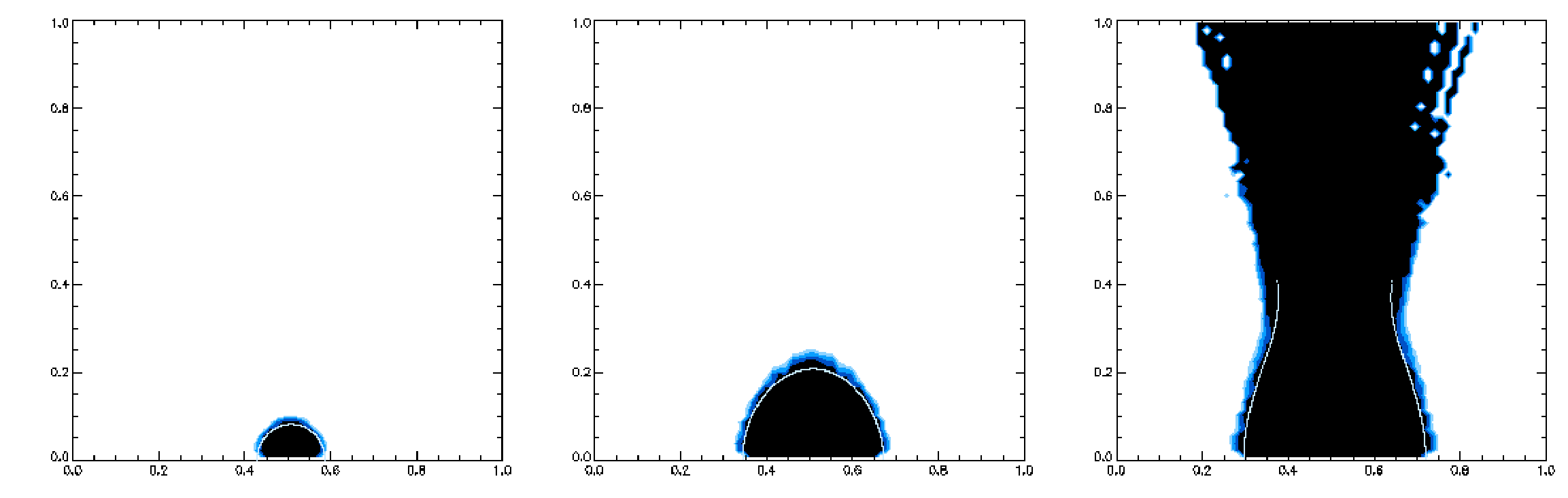,height=5.5cm}
\caption{\label{fig02}\footnotesize{Evolution of ionised regions (black)
produced in the GDD described in Section~2, by sources with
different ionisation rate, located on the midplane of the disk.
The linear box size is 1~kpc. Columns, from the left to the right,
refer to a source ionisation rate of $\dot{\cal N}_\gamma=
(1.8, 16, 34) \times 10^{48}$~s$^{-1}$
respectively; from the top to the bottom the ionisation structure at a time
$t=(0.05, 1, 1.5, 5)$~Myr after source turn on is shown.
The solid lines correspond to the stationary analytical solution
(Dove \& Shull 1994b).}} 
\end{figure*}

To study the propagation of ionising radiation produced by the
stars through the given ISM density distribution we use the
Monte Carlo (MC) radiative transfer code {\tt CRASH} ({\it Cosmological
RAdiative transfer Scheme for Hydrodynamics}), described in Ciardi \etal
(2001). The code has been originally developed for cosmological
applications, but below we show that it can correctly handle also 
the rather different conditions prevailing in the interstellar medium of
galaxies. For clarity, we briefly summarize the main features of the
numerical scheme relevant to the present study.

In the application of a MC scheme to radiative transfer problems, the
radiation intensity is discretized into a representative number of
monochromatic photon packets. The processes involved (e.g. packet
emission and absorption) are then treated statistically by randomly
sampling the appropriate distribution function. 
The 1 kpc$^3$ simulation volume has been discretized in $64^3$ grid cells;
we consider absorption by HI and dust; we neglect
the contribution of He because of the paucity of the He-ionising photons
produced.  
We have estimated the dust optical depth in the optical using 
the canonical relation between the HI column density, $N_{HI}$, and the 
color excess E(B-V) (Bohlin, Savage \& Drake 1978). This has 
been converted into an  optical depth at the ionisation limit 
by adopting the far-UV parametrization of the mean Galactic 
extinction law (Fitzpatrick \&Massa 1988). We obtain:
\be
\tau_d(912\AA)=1.9 \times 10^{-21} (N_H/ {\rm cm}^{-2}),
\ee
where we have excluded the contribution of dust scattering to the
opacity, by multiplying for $(1-\omega)$; $\omega=0.4$ is the average 
albedo of dust in the UV close to the ionisation limit (Witt \etal 1993).
Given the above equation, as the final mean ionisation fraction for 
run A (run B) is $\langle x \rangle =0.69$ (0.40) (see Section~4), for the 
frequencies of interest here, we find that $\tau_d/\tau_{HI}<2.46 \times 10^{-2}$
($1.28 \times 10^{-2}$). Thus, dust contribution to absorption is negligible.

For the range of densities considered here, a number of photon
packets equal to ${\cal N}_p=5 \times 10^7$ has to be emitted in
order to reach numerical convergence. 

Differently from Ciardi {\it et al.} (2001), here we deal with
more than one ionising source. We have found that, if the
number of emitted packets is sufficiently high, a 
more reliable solution
of the discretized time-dependent ionisation equation is
obtained if the time step $\Delta t$ used in their eq.~10 is
not treated statistically. Thus, here $\Delta t$ is the time
elapsed since a photon packet has gone through the cell for which
the equation is being solved.
Only photons escaping from the top side of the cubic region 
contribute to the escape fraction, while those escaping from side faces
or the bottom one are subject to a reflecting boundary condition, thus simulating photons
coming from adjacent regions. 

The same numerical tests of the code described in Ciardi {\it et al.} (2001)
have been performed in an interstellar (rather than intergalactic) environment.
We do not report them here; instead,
in Fig.~\ref{fig02} we show the evolution of ionised regions (in black) 
produced in the GDD
described in the previous Section, by sources with different ionisation
rate, located on the disk midplane. The slices show   
a plane perpendicular to the Galactic disk through each source location. 
Different columns refer to different source ionisation rates 
($\dot{\cal N}_\gamma=1.8, 16, 34 \times 10^{48}$~s$^{-1}$ respectively, 
from left to right), while
from the top to the bottom we show the ISM ionisation structure at times 
$t=0.05, 1, 1.5, 5$~Myr after the source has been turned on. The solid
lines correspond to the stationary analytical 
solution given in Dove \& Shull (1994b).
The ionisation rates were chosen in order to reproduce the various
possible shapes of the ionised regions: spherical, elongated and funnel-like.

To prevent misunderstandings, it is useful here to give 
our working definition of the escape fraction. We define
the {\it global} escape fraction, $f_{esc,g}$, as: 
\be
f_{esc,g}(t) \equiv {\int_0^{t} dt \dot {\cal N}_\gamma^e(t) \over 
\int_0^{t} dt \dot {\cal N}_\gamma(t)},
\label{fglob}
\ee
where $\dot {\cal N}_\gamma^e$ is the escaping photon rate ({\it i.e.} 
the number of photons reaching $z \ge 1$~kpc per unit time) 
and $\dot{{\cal N}}_\gamma$ is the total photon rate
production at (simulation) time $t$.
The {\it instantaneous} escape 
fraction, $f_{esc,i}$, is instead given by: 
\be
f_{esc,i}(t) \equiv {\dot {\cal N}_\gamma^e(t) \over 
\dot {\cal N}_\gamma(t)}.
\label{finst}
\ee

In order to derive $f_{esc}$, we assume that the stars have a 
constant ionisation rate and we
run the simulation until convergence (\ie the escape fraction
and the ionisation structure do not change) is reached. 
As we find that, after a time of $\sim 5 \times 10^6$~yr the value of 
$f_{esc}$ remains roughly constant, we stop the simulations after 
$\sim 10^7$~yr. This is expected as the recombination 
time at the midplane is 0.3 Myr. 

Following the method described above, we have run 20 simulations 
for each of the cases indicated in Table~1.

\section{Results}

\begin{figure}
\psfig{figure=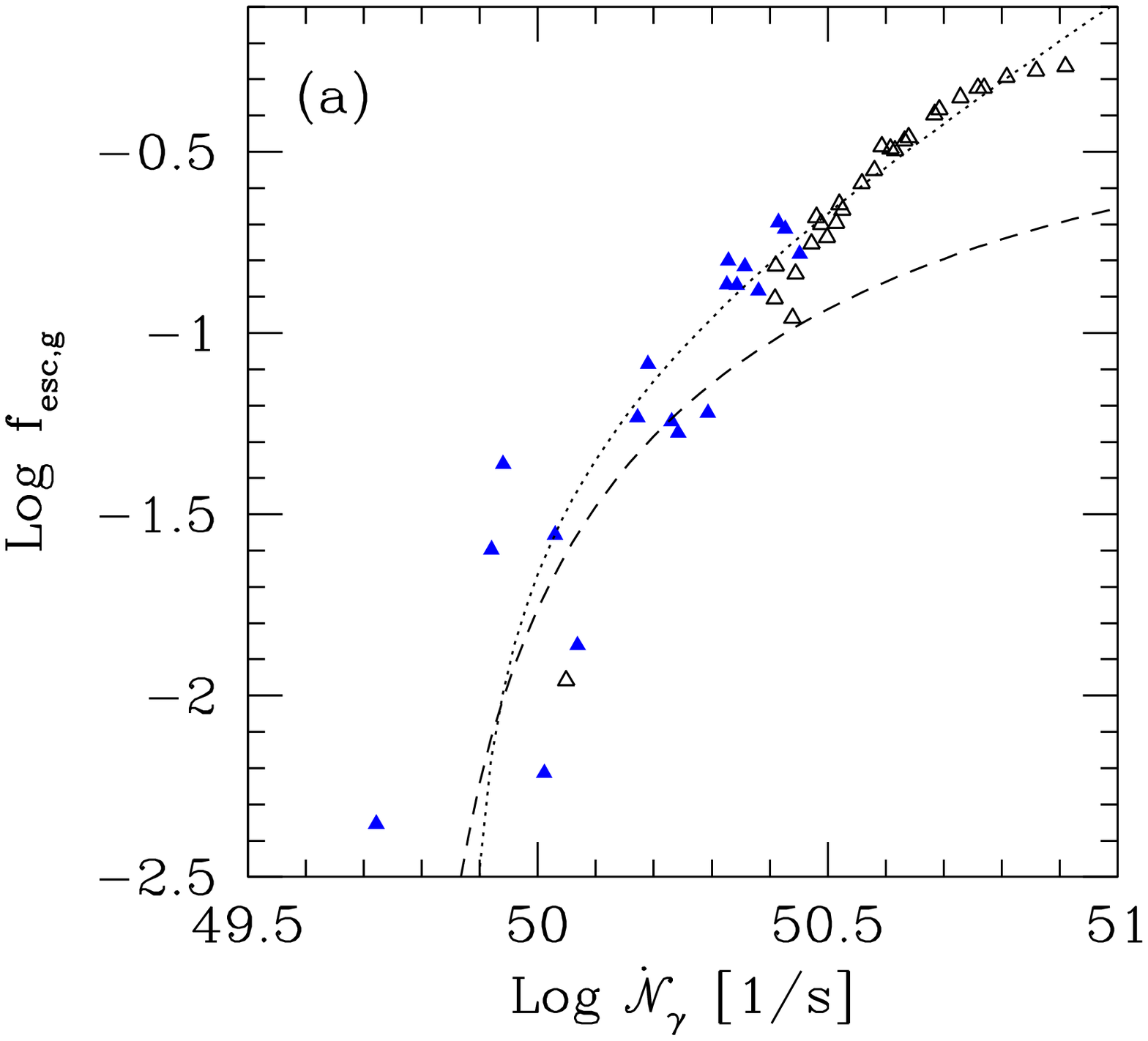,height=7.5cm}
\vskip 0.5truecm
\psfig{figure=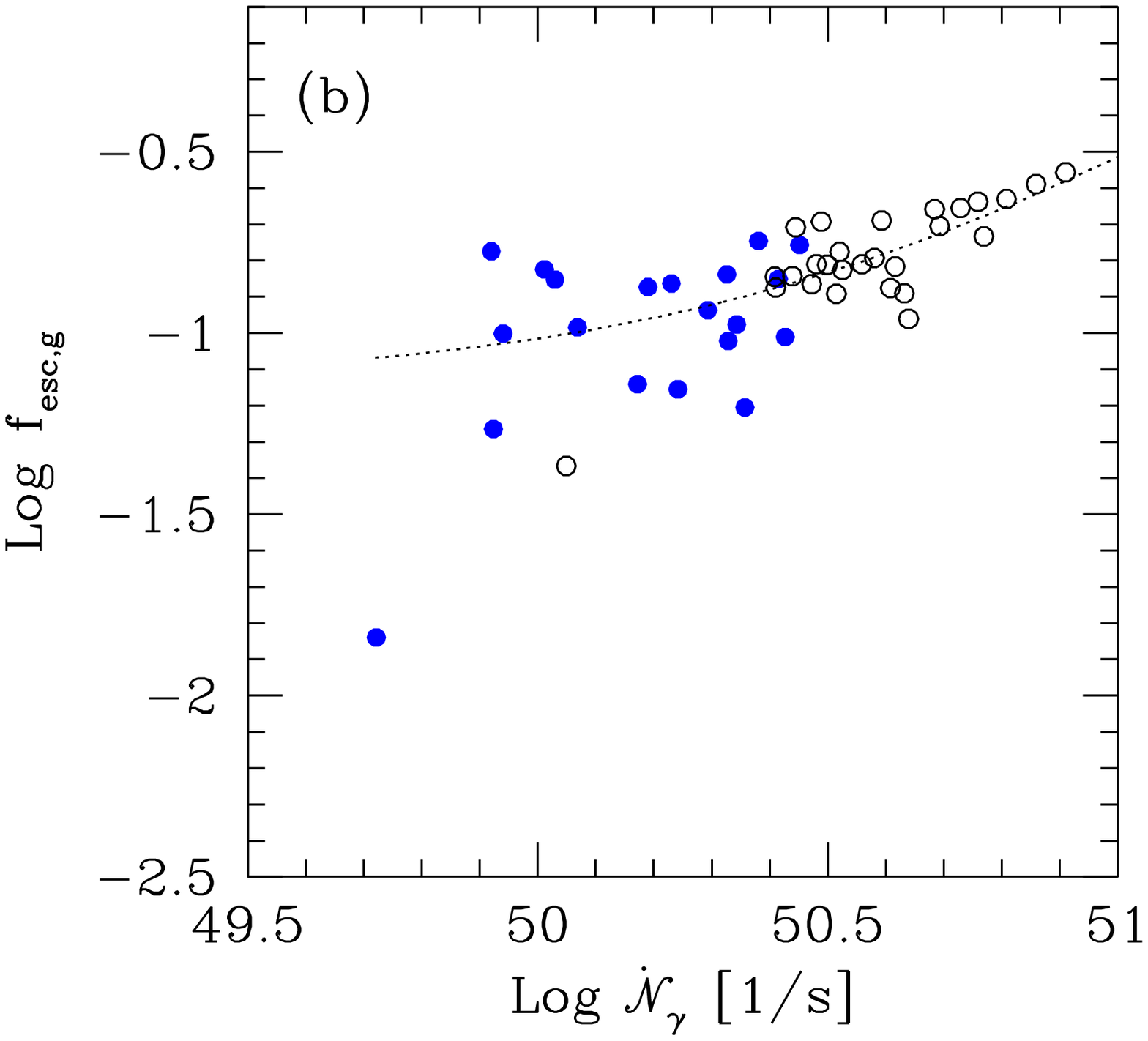,height=7.5cm}
\caption{\label{fig03}\footnotesize{(a) Evolution of the global escape
fraction, $f_{esc,g}$, as a function of the total ionisation rate,
$\dot{{\cal N}}_\gamma$, for runs A (filled triangles) and C
(open triangles). The dotted line is the fit given by eq.~\ref{fitds}. 
The dashed line is the analytical solution derived by Dove \& Shull
(1994a), assuming a single source located on the galactic plane. 
(b)  Same as (a) for runs B (filled circles) and D (open circles).}}
\end{figure}

In Fig.~\ref{fig03} we show the evolution of the global escape
fraction, $f_{esc,g}$, as a function of
the total ionisation rate, $\dot{{\cal N}}_\gamma$, for 
runs A (filled triangles in Fig.~\ref{fig03}a)
and B (filled circles in Fig.~\ref{fig03}b). 
In both cases $f_{esc,g}$ increases with increasing
$\dot{{\cal N}}_\gamma$, although for the FDD with a flatter slope.
The reason for this different behavior is that photons in a GDD can escape
only if an ionised channel can be produced by the stars themselves. 
Consequently, a low ionisation rate results in a low escape fraction.
In a FDD instead, photons can travel along clear sigthlines through low 
density ionised channels and escape more easily. Thus, for 
$\dot{{\cal N}}_\gamma \simlt 2 \times 10^{50}$~s$^{-1}$
$f_{esc,g}$ is always higher for a FDD, while for larger ionisation rates
a GDD becomes more transparent. In both cases, the scatter in $f_{esc,g}$ is
mainly due to differences in the star position and ionisation rate.

To study the effect of a larger total ionisation rate, we have increased
the stellar surface density, as explained in Section~2, obtaining a
maximum value $\dot{{\cal N}}_\gamma \simeq 5  \times 10^{50}$~s$^{-1}$. 
These additional runs are represented in Fig.~\ref{fig03} as open   
triangles (run C) and circles (run D). The scatter is reduced with respect to 
runs A and B as, increasing the number of stars, their position affects
less sensibly the value of $f_{esc}$, which instead depends
more strongly on the total ionisation rate. In order to check if the curve for the 
GDD would reach a plateau or rather keep growing (as the analytical
derivation by Dove \& Shull 1994a suggests), we have increased
the total ionisation rate up to a value of $\simeq 8 \times 10^{50}$~s$^{-1}$
(corresponding to a stellar surface density of 
$\Sigma_\star \simeq 120$~kpc$^{-2}$). 
As we expect the scatter to be reduced, in order to
limit the computational time, we have only run 6 different realizations.

The evolution of $f_{esc,g}$ is linear over the considered range of 
$\dot{{\cal N}}_\gamma$ and it can be fitted by the
following function:
\be
\label{fitds}
f_{esc,g}=\alpha (\dot{{\cal N}}_\gamma/10^{50} {\rm s}^{-1})+\beta,
\ee
where for a GDD (FDD) $\alpha=0.089 \pm 0.003$ ($0.023 \pm 0.003$)
and $\beta=-0.07 \pm 0.01$ ($0.07 \pm 0.01$).
In both cases the fit is represented as a dotted line in Fig.~\ref{fig03}.
We find that the GDD curve eventually flattens and departs from the fit 
in eq.~\ref{fitds}; on the contrary the results for the FDD are well
described by the same fit also for high values of $\dot{{\cal N}}_\gamma$.

Dove \& Shull (1994a) derived analytically the value of $f_{esc}$ expected in a
GDD from a point source of given ionisation rate, located on the 
disk of the Galaxy. The value they obtained (represented by a dashed line
in Fig.~\ref{fig03}a) should be compared with the asymptotic value
of $f_{esc,i}$. However, as for late evolutionary times $f_{esc,g}$ approaches
$f_{esc,i}$ (see below), we can compare their results with our runs 
A and C directly from Fig.~\ref{fig03}a. Clearly, the evolution of $f_{esc,g}$ 
differs from the analytic calculation and higher values for the escape
fraction are obtained, especially for large ionisation rates. This is due to the
fact that in our case the ionisation rate is distributed among various
sources located at different heights above the midplane.

\begin{figure}
\vskip -1.truecm
\psfig{figure=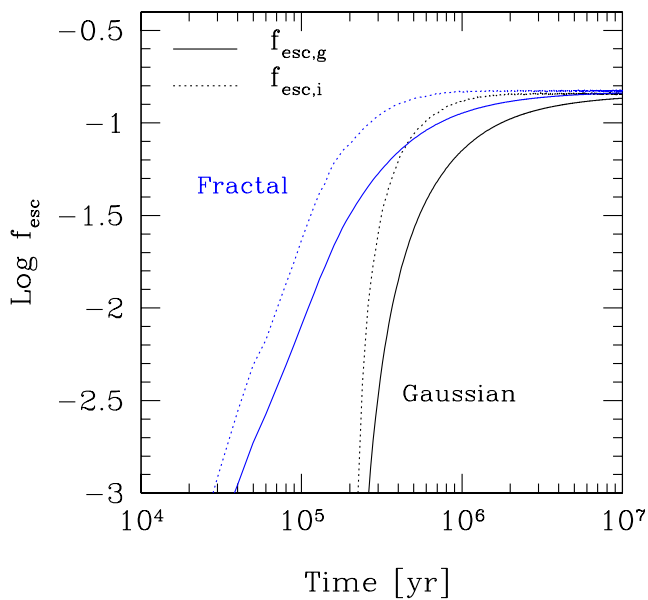,height=10cm}
\caption{\label{fig04}\footnotesize{Time evolution of the escape fraction 
for a Gaussian (lower curves) and a fractal (upper curves) density 
distribution. The evolution of $f_{esc,g}$ (solid lines), as well as
the one of $f_{esc,i}$ (dotted lines) is shown.}}
\end{figure}

In Fig.~\ref{fig04} the time evolution of the escape fraction is shown,
for a Gaussian (lower curves) and  a fractal (upper curves) density
distribution. As a reference, we have chosen a run that produces, for
both distributions and given
the same set of stars (with $\dot{{\cal N}}_\gamma= 
2 \times 10^{50}$~s$^{-1}$), a comparable value of $f_{esc,g}$.
The solid (dotted) lines indicate the evolution of $f_{esc,g}$
($f_{esc,i}$). As in a GDD photons can escape only through 
ionised channels produced by the stars themselves,
the photon escape is retarded and the evolution of 
$f_{esc}$ is steeper compared to the case of a FDD.             
Initially, $f_{esc,i}$, which becomes constant
after $\approx  10^6$~yr, is higher than $f_{esc,g}$.
This is due to the fact that, when the first photons escape (at a time
$t \sim  10^3 [10^5]$~yr for a FDD [GDD]), a large number of
photons has already been emitted and this results in a very
low value of $f_{esc,g}$. Then $f_{esc,g}$ increases, approaching
$f_{esc,i}$. 

\begin{figure*}
\psfig{figure=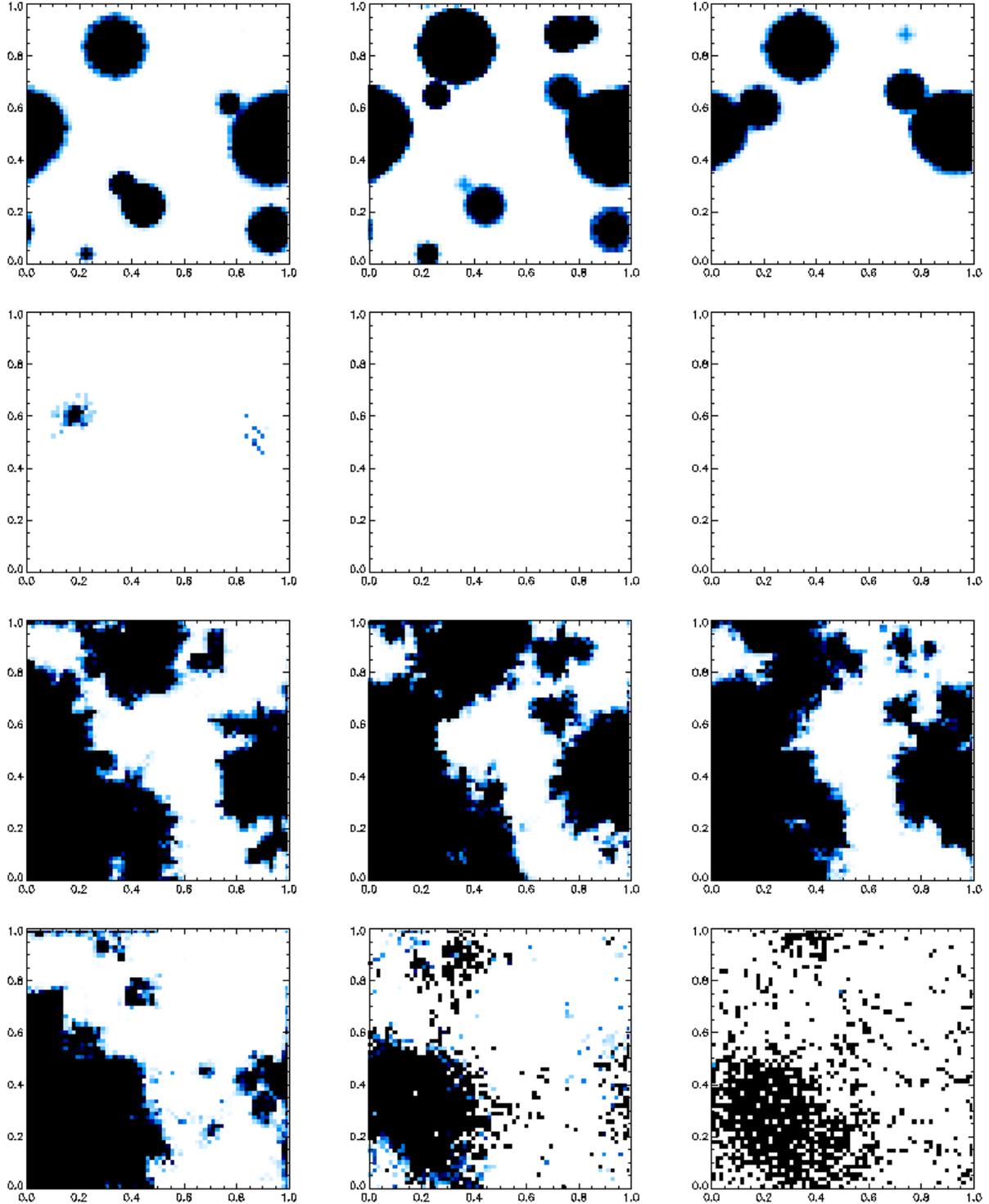,height=20cm}
\caption{\label{fig05}\footnotesize{Evolution of ionisation
field (black) produced by a given stellar distribution into a
Gaussian (top six panels) and a fractal (bottom six panels)
density field. The slices are taken at a time $t=0.1$~Myr
from the source turn on, at different vertical distances
from the midplane (0, 10, 156, 312, 625, 940~pc, left to right).}}
\end{figure*}

\begin{figure*}
\psfig{figure=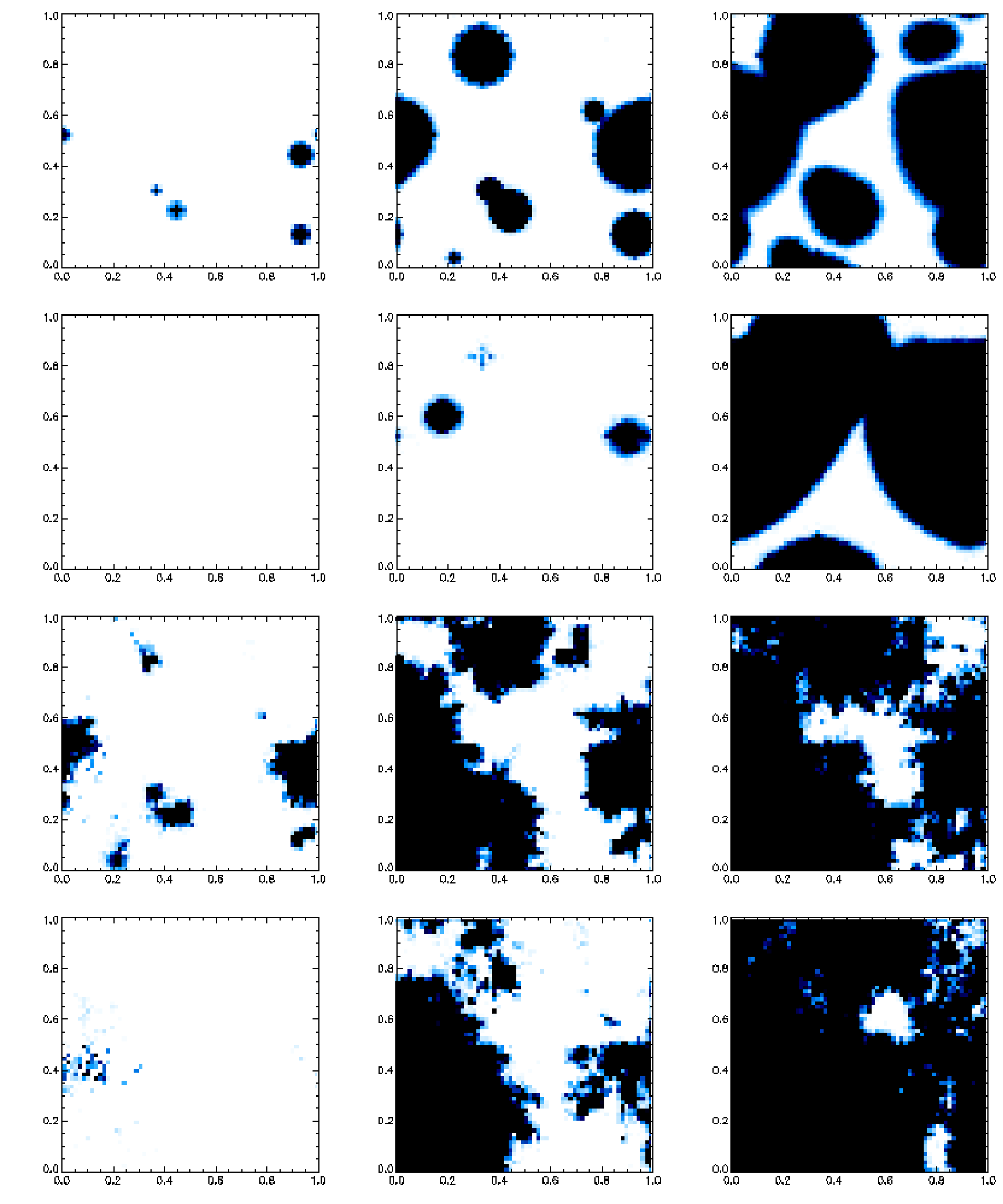,height=20cm}
\caption{\label{fig06}\footnotesize{Evolution of ionisation
field   (black) produced by a given stellar distribution into a
Gaussian (top six panels) and a fractal (bottom six panels)
density field. The time evolution (columns, from left to right,
refer to $t=0.005, 0.5, 5$~Myr respectively) of slices at 0 
(first row) and 265~pc (second row) above the midplane is shown.}}
\end{figure*}

In Figs.~\ref{fig05} and~\ref{fig06} illustrative slices
extracted from the simulation box for the same run of Fig.~\ref{fig04},
show the evolution of the ionised gas (black regions) produced by the 
sources in the given density field. The top six panels are
for a GDD and show the ionised regions at
a time $t=0.1$~Myr from the source turn on, at different vertical
distances from the midplane.
The bottom six panels are the same for a FDD. 
As already pointed out above, although in the GDD
the upper gas layers at this time are still neutral, some ionising photons
have already escaped the FDD.           
In Fig.~\ref{fig06} the top (bottom) six panels refer to the GDD (FDD).
Here we show the time evolution 
(columns, from left to right, refer to $t=0.005, 0.5, 5$~Myr respectively)
of slices at two heights above the midplane.
Again, the ionised regions are more extended in a fractal than in
a Gaussian medium.

Although in this work we focus on the variation of $f_{esc}$ with the
different density distribution, it is interesting to check the values we
obtain for the column density of the neutral gas. For runs A and B, we 
should be able to recover the observed Galactic HI column density 
($N_{HI}=5.2 \times 10^{20}$~cm$^{-2}$), i.e. about 30\% of the total gas
should be ionised (Sect.~\ref{ismden}). For a GDD (run A) we obtain a final 
mean HI column density of $N_{HI} =  2.1 \times 10^{20}$~cm$^{-2}$,
\ie about 40\% lower; for a FDD (run B), $N_{HI} = 4.1 \times 
10^{20}$~cm$^{-2}$ is closer to the observed value. In both cases, 
however, some of the simulations have column density closer to the
observed.  Despite the larger escape fraction for a FDD, the resulting 
mean HI column density is larger for a fractal medium, where the highest
density regions are more difficult to ionise and recombine faster. 
Also with an increased stellar surface density (runs C and D)
the mean $N_{HI}$ in a FDD remains as high as 
$\approx 2.7 \times 10^{20}$~cm$^{-2}$, while in a GDD it is $\approx 5.9 
\times 10^{19}$~cm$^{-2}$, with values as low as $\approx 10^{17}$~cm$^{-2}$.
The reduced ionisation fraction in the case of an inhomogeneous medium 
was also noted in the model for the diffused ionised medium of Miller \& Cox 
(1993), which took into account the absorption of UV radiation by clouds with 
density larger than the surrounding medium in a statistical way.

\begin{figure}
\psfig{figure=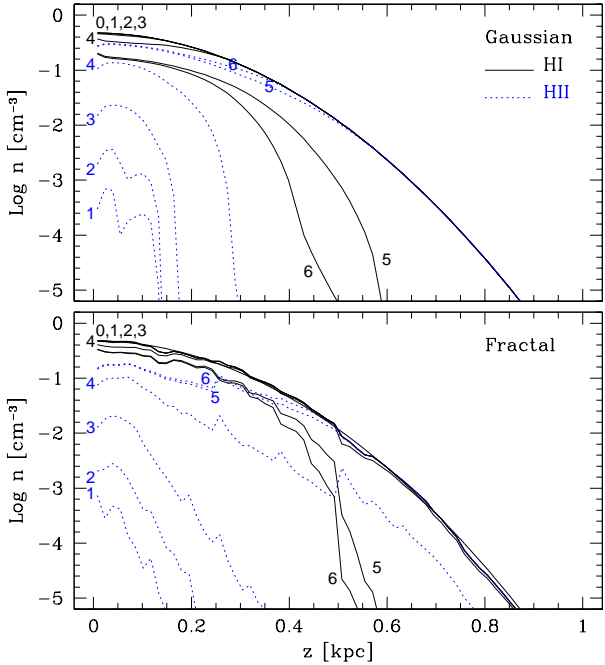,height=9.5cm}
\caption{\label{fig07}\footnotesize{Vertical distribution of the mean HI
(solid lines) and HII (dotted lines) number densities, in the
case of a Gaussian (top panel) and fractal (bottom panel) density
field. The numbers refer to different times from the source turn on:
$i=0...6$ refers, respectively, to
$t=0, 10^2, 10^3, 10^4, 10^5, 10^6$ and $10^7$~yr.}}
\end{figure}

In Fig.~\ref{fig07} the vertical distribution of the mean HI (solid lines) 
and HII (dotted lines) number densities is shown, for runs A and B with the 
above $N_{HI}$, in the case of a Gaussian (top panel) and fractal (bottom 
panel) density field at different times after the source turn on.
For both distributions, the HI profile does not change until  
$t \approx 10^5$~yr, when the inner regions, where the stars reside,
have an HII density of $n_{HII} \approx 0.1$~cm$^{-3}$. 
The final HI density profile has a similar shape for both 
distributions, with a slightly lower value in the inner regions for
a GDD case. The HII number density increases gradually in a FDD, while
in a GDD, as previously mentioned, the ionisation of the upper gas
layers is postponed until the stars have been able to produce an ionised 
channel for the escaping of photons. At the final stages, the HII 
distribution appears more extended than the HI one. This is consistent 
with observations of the Reynolds layer, which seems to have a vertical 
scale height of $\approx 1$~kpc (Mezger 1978; Reynolds 1991a, 1991b). 
However, a proper modelling of the diffuse ionised gas should take 
into account the dynamical effects on gas, which would make the ionised gas 
to expand, as well as a larger vertical size for the simulation box.

\begin{figure}
\vskip -1.truecm
\psfig{figure=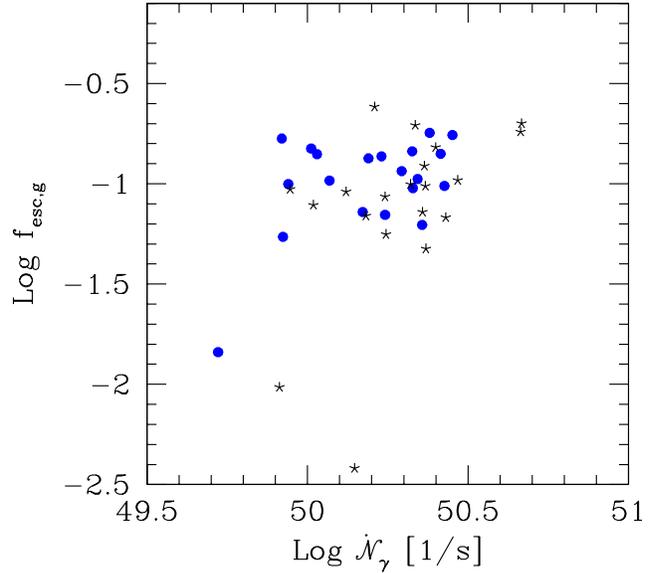,height=10cm}
\caption{\label{fig08}\footnotesize{Evolution of the global escape
fraction, $f_{esc,g}$, as a function of the total ionisation rate,
$\dot{{\cal N}}_\gamma$, for runs B (filled circles) and E
(stars).}}
\end{figure}

In Fig.~\ref{fig08} we show the evolution of $f_{esc,g}$ as a function of
the total ionisation rate for run B (filled circles) and E (stars), the
only difference being the stellar distribution. The distribution of the escape 
fraction looks similar in the two cases and the obtained values are
comparable: in fact, in run E the stars, following the Gaussian distribution 
of the gas, have a larger vertical extension, but, as they are located in 
denser regions, more photons are needed to ionised them and the two 
effects roughly balance.

\begin{figure*}
\psfig{figure=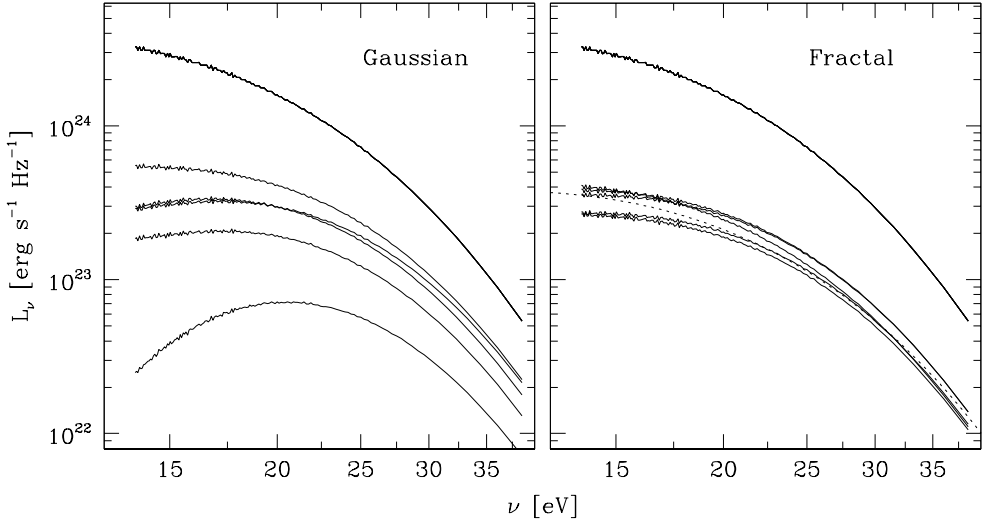,width=13cm}
\caption{\label{fig09}\footnotesize{Spectrum of the emerging ionising
radiation, for five realizations of run A (left panel) and run B
(rigth panel). In both panels, the uppermost curve is the spectrum 
of the intrinsic ionising radiation (a blackbody with T=40000~K). For
ease of presentation, all the spectra have been normalised to an
intrinsic photon rate $\dot{{\cal N}}_\gamma = 2\times 10^{50} \mbox{s}^{-1}$.
The dotted curve in the right panel is a blackbody with T=45000~K.
}}
\end{figure*}

Finally, we present in Fig.~\ref{fig09} the spectrum of the emerging
ionising radiation, for five realizations of runs A and B, respectively. 
In both plots,
the uppermost curve is the intrinsic spectrum of stars, i.e. a black 
body with T=40000~K. As already discussed, each simulation is characterised 
by the rate of emitted ionising photons, $\dot{{\cal N}}_\gamma$. Since
we are mainly interested in the shape of the spectrum, we normalise all
simulations to the value $\dot{{\cal N}}_\gamma = 2\times 10^{50}
\mbox{s}^{-1}$. As expected from the $\nu^{-3}$ frequency dependence 
of the HI photoionisation cross section, the emerging filtered spectrum 
is harder than the intrinsic one. For run A, the shape of the spectrum 
depends on the amount of escaping photons (\ie on the escape fraction) 
in each realization: the lower and upper emerging spectra in the left 
panel of Fig.~\ref{fig09} being those for the lower and higher values
of $\dot{{\cal N}}_\gamma$. Different realizations of run B yield much
more similar  emerging spectra, this reflecting the nearly constant value of
the escape fraction. As a reference, we plot in the right panel of 
Fig.~\ref{fig09}
a blackbody with T=45000~K. We remind that here we do not take into
account the absorption of photons with $h\nu>24.6$ eV by HeI, which 
may be the cause for the low HeI ionisation observed in the diffuse
ionised medium (Reynolds \& Tufte 1995).

\section{Summary and Conclusions}

We have studied the effects of gas density inhomogeneities on the 
escape of ionizing Lyman continuum photons from Milky Way-type galaxies
via radiative transfer numerical simulations. To this aim a comparison
between a smoothly stratified and an inhomogeneous, fractal distribution
have been made with realistic assumptions for the ionizing stellar sources
based on available data in the solar neighborhood. 
The main results obtained can be summarized as follows.

(1) The global escape fraction, $f_{esc,g}$, in the case  
of a Gaussian Density Distribution (GDD) rapidly increases with
increasing total ionization rate, $\dot{{\cal N}}_\gamma$, and it 
flattens for high values of $\dot{{\cal N}}_\gamma$; for a
Fractal Density Distribution (FDD) the dependence on 
$\dot{{\cal N}}_\gamma$ is milder.

(2) For $\dot{{\cal N}}_\gamma \simlt 2 \times 10^{50}$~s$^{-1}$
$f_{esc,g}$ is always higher for a FDD, while for larger ionization rate
a GDD becomes more transparent. 

(3) For low stellar surface densities, $\Sigma_\star$, there is a large 
scatter in the results, due to differences in the star positions and 
ionization rates. The scatter is reduced for larger $\Sigma_\star$, as the
star position affects less sensibly the value of $f_{esc}$, which 
instead depends more strongly on $\dot{{\cal N}}_\gamma$.

(4) In a GDD the photon escape is retarded and the time evolution of
the escape fraction is steeper compared to the FDD case.

(5) For a GDD we obtain a final mean HI
column density of $N_{HI} = 2.1 \times 10^{20}$~cm$^{-2}$,
lower than the observed one;
for a FDD, $N_{HI} = 4.1 \times 10^{20}$~cm$^{-2}$ closely matches 
the experimental value. In all cases, the HII distribution
appears to be more extended than the HI one.

(6) The value of the escape fraction is more sensitive to the
gas density rather than the stellar distribution.

In order to draw a more general conclusion from the results summarized
above, it is useful to translate the total ionization rate into a
star formation rate for a given galaxy. To do so we use the results of the
stellar population synthesis code Starburst99 (Leitherer \etal 1999). 
Assuming a continuous star formation mode, a Salpeter IMF with $M_{down}
= 1 M_\odot$ and $M_{up}= 100 M_\odot$ and a metallicity equal to 1/20 of
solar, we find that the expected ionization rate inside our simulated 
kpc$^3$ produced by the corresponding stellar population is $\dot{{\cal N}}_\gamma =
4.5\times 10^{50} (\dot M_\star/M_\odot {\rm yr}^{-1})$. From Fig. 
\ref{fig03} we then note that for $\dot M_\star \approx 2 M_\odot {\rm
yr}^{-1}$ the escape fraction is already equal to roughly 0.6 (0.28)
for the GDD (FDD) case. 

Star formation rates of a few tens of solar masses are commonly derived from
observations of LBGs (Pettini \etal 2001).  This result may suggest a
galaxy-dominated UV background. In fact, Bianchi, Cristiani \& Kim (2001)
have recently shown that estimates of the local and high-$z$ meta-galactic
ionizing flux are consistent with a galaxy-dominated background if 
$f_{esc}\approx 10\%$. Hence it appears that these
type of sources, as long as they could be modelled as relatively normal disk
galaxies, can strongly influence the ionization history of the intergalactic
medium and, possibly, the galaxy formation process. 

In principle, then, photons can escape rather efficiently (particularly if 
their ISM is smooth) from high redshift galaxies, 
provided their vertical gas distribution is not too different from the
Milky-Way type galaxies investigated here. Although galactic disk models
based on simple semi-analytical prescriptions (Fall \& Efstathiou 1980;
Mo, Mao \& White 1998; Weil, Eke \& Efstathiou 1998; Ferrara, Pettini
\& Shchekinov 2000; Wood \& Loeb 2000) predict that the disk 
column density should increase with redshift by a factor $\approx 10$
between the present value and redshift 5, this estimate is still 
subject to several uncertainties. For example it is not clear how both
the gas temperature (governing the disk scaleheight) and the fraction
of baryons that are able to cool and settle in the disk (determining
the midplane gas density) evolve with cosmic time. Even if the column 
density is larger, we do expect that the galactic ISM is smoother  
than the present day one, due to the short 
time interval available for the build up of a fully developed turbulent
spectrum and hence a fractal gas distribution. This process might
require several (say, 10) eddy turnover times, $t_e$, on the scale at 
which turbulence is injected by superbubbles, \ie $ \ell \approx 1$~kpc 
(Norman \& Ferrara 1996).
If the characteristic eddy velocity is $v_t \approx 1 $~km~s$^{-1}$, this 
implies that $t_e \approx 10$~Gyr is much longer than the Hubble time
at redshift 3. 
Hence,
a final conclusion from  models like the one presented
here extended to encompass high redshift galaxies have to await 
additional data (both observational and from galaxy formation models) 
on the distribution and dynamical state of the gas in primordial
galaxies.

\section*{Acknowledgments}

We would like to thank A. Lorenzani and M. Normandeau for help with 
the HI density distributions and the referee, K. Wood, for useful
comments. 
This work was partially supported by the Research and Training Network 
"The Physics of the Intergalactic Medium" set up by the European Community 
under the contract HPRN-CT2000-00126 RG29185.

\label{lastpage}

\newpage


\begin{thebibliography}{99}

\bibitem{BNSL01} Benson, A. J., Nusser, A.,  Sugiyama, S. \& Lacey, C.
G. 2001, MNRAS, 320, 153                

\bibitem{BCK01} Bianchi, S., Cristiani, S. \& Kim, T.-S. 2001, A\&A, 376, 1   

\bibitem{BHM99} Bland-Hawthorn, J. \&  Maloney, P. R. 1999, ApJ, 510,
33

\bibitem{BHM01} Bland-Hawthorn, J. \&  Maloney, P. R. 2001, ApJ, 550,
231

\bibitem{BSD78} Bohlin, R. C., Savage, B. D. \& Drake, J. F. 1978, ApJ, 224, 132

\bibitem{BH94} Burton, W. B. \& Hartmann, D. 1994, Ap\&SS, 217, 189

\bibitem{CFGJ00} Ciardi, B., Ferrara, A., Governato, F. \& Jenkins, A.
2000, MNRAS, 314, 611                      

\bibitem{CFMR01} Ciardi, B., Ferrara, A., Marri, S. \& Raimondo, G. 2001,
MNRAS, 324, 381                 

\bibitem{DL90} Dickey, J. M. \& Lockman, F. J. L. 1990, ARA\&A, 28, 215

\bibitem{DS94a} Dove, J. B. \& Shull, J. M. 1994a, ApJ, 423, 196

\bibitem{DS94b} Dove, J. B. \& Shull, J. M. 1994b, ApJ, 430, 222

\bibitem{DSF00} Dove, J. B., Shull, J. M. \& Ferrara, A. 2000, ApJ, 531, 846

\bibitem{E97} Elmegreen, B. G. 1997, ApJ, 477, 196

\bibitem{E99} Elmegreen, B. G. 1999, ApJ, 527, 266 

\bibitem{EF96} Elmegreen, B. G. \& Falgarone, E. 1996, ApJ, 471, 816

\bibitem{FPW91} Falgarone, E., Phillips, T. G. \& Walker, C. K. 1991, ApJ, 378, 186

\bibitem{Fa80} Fall, S. M. N. \& Efstathiou, G. 1980, MNRAS, 193, 189

\bibitem{Fe00} Ferrara, A., Pettini, M., \& Shchekinov, Y. 2000, MNRAS, 319, 539

\bibitem{FM88} Fitzpatrick, E. L. \& Massa, D. 1988, ApJ, 328, 734

\bibitem{GFM97} Giallongo, E., Fontana, A. \& Madau, P. 1997, MNRAS, 289, 629

\bibitem{GCC82} Garmany, C. D. , Conti, P. S.  \& Chiosi, C. 1982, 263, 777

\bibitem{GS97} Giroux, M. \& Shull, J. M. 1997, AJ, 113, 1505

\bibitem{GO98} Gnedin, N. Y. \& Ostriker, J. P. 1998, ApJ, 486, 581

\bibitem{G00} Gnedin, N. Y. 2000, ApJ, 535, 530                  

\bibitem{HMKH01} Haehnelt, M. G., Madau, P., Kudritzki, R. \&
Haardt, F. 2001, ApJL, 549, 151

\bibitem{H01} Heckman, T. M. \etal 2001, ApJ, 558, 56

\bibitem{HJD97} Hurwitz, M., Jelinsky, P. \& Dixon, W. V. 1997, ApJ, 481, L31

\bibitem{KF85} Kulkarni, S. R. \& Fich, M. 1985, ApJ, 289, 792


\bibitem{LFHL95} Leitherer, C., Ferguson, H. C., Heckman, T. M. \&
Lowental, J. D. 1995, ApJL, 454, 19

\bibitem{L99} Leitherer, C. \etal  1999, ApJS, 123, 3

\bibitem{MA01} Ma\'{\i}z-Apell\'aniz, J. 2001, AJ, 121, 2737                       
\bibitem{M78} Mezger, P. G. 1978, A\&A, 70, 565                        

\bibitem{M93} Miller, W. W. \& Cox, D. P. 1993, ApJ, 417, 579. 

\bibitem{MEHR00} Miralda-Escud\'e, J., Haehnelt, M. \& Rees, M. R.
2000, ApJ, 530, 1

\bibitem{M98} Mo, H. J., Mao, S. \& White, S. D. M. 1998, MNRAS, 295, 319

\bibitem{NF96} Norman, C. A. \& Ferrara, A. 1996, ApJ, 467, 280 

\bibitem{P01} Pettini, M. \etal 2001, ApJ, 554, 981

\bibitem{R85} Reynolds, R. J. 1985, ApJ, 294, 256

\bibitem{R91a} Reynolds, R. J. 1991a, ApJ, 372, L17

\bibitem{R91b} Reynolds, R. J. 1991b, in The Interstellar Disk-Halo
Connection in Galaxies, ed. H. Bloemen, (Dordrecht: Kluwer), p. 67

\bibitem{R93} Reynolds, R. J. 1993, Massive Stars: Their Lives in
the Interstellar Medium, ASP Conference Series, eds. J.P.
Cassinelli \& E.B. Churchwell, vol. 35, p. 338

\bibitem{RH01} Reynolds, R. J. \& Haffner, L. M. 2001, preprint (astro-ph/0010618)

\bibitem{RH95} Reynolds, R. J. \& Tufte, S. L. 1995, ApJ, 439, L17.

\bibitem{RS00} Ricotti, M. \& Shull, J. M. 2000, ApJ, 542, 548  

\bibitem{SKNF77} Sanders, B. D., Kraushaar, W. L., Nousek, J. A. \&
Fried, P. M. 1977, ApJ, 217, L87

\bibitem{SdK97} Schaerer, D. \& de Koter, A. 1997, A\&A, 322, 598

\bibitem{SLCW99} Sfeir, D. M., Lallement, R., Crifo, F. \& Welch, B. Y.
1999, A\&A, 346, 785

\bibitem{SCMS90} Snowden, S. L., Cox, D. P., McCammon, D. \& Sanders,
W. T. 1990, ApJ, 354, 211

\bibitem{SPA01} Steidel, C. C., Pettini, M. \& Adelberger, K. L. 2001
ApJ, 546, 665 

\bibitem{W98} Weil, M. L., Eke, V. R. \& Efstathiou, G. 1998, MNRAS, 300, 773

\bibitem{W93} Witt, A. N. \etal 1993, ApJ 410 714

\bibitem{WL00} Wood, K. \& Loeb, A. 2000, ApJ, 545, 86
\end{thebibliography}
\end{document}